\documentclass[runningheads]{llncs}

\usepackage[T1]{fontenc}
\usepackage{graphicx}
\usepackage{amsmath}
\usepackage{xcolor}
\usepackage{amssymb}
\usepackage{bbding}
\usepackage[linesnumbered,lined,ruled,noend]{algorithm2e}
\usepackage{booktabs}
\usepackage{multirow}
\usepackage{wrapfig}

\begin{document}

\title{Strategic Sample Selection for Improved Clean-Label Backdoor Attacks in Text Classification}

\titlerunning{Strategic Sample Selection for Clean-Label Backdoors}

\author{Onur Alp Kirci\orcidID{0009-0004-0230-1453} \and M.~Emre~Gursoy\Envelope\orcidID{0000-0002-7676-0167}\thanks{Corresponding author: emregursoy@ku.edu.tr}}

\authorrunning{Kirci and Gursoy}

\institute{Department of Computer Engineering, Koç University, Istanbul, Turkey \\ \email{\{okirci21, emregursoy\}@ku.edu.tr}}

\maketitle  

\begin{abstract}
Backdoor attacks pose a significant threat to the integrity of text classification models used in natural language processing. While several dirty-label attacks that achieve high attack success rates (ASR) have been proposed, clean-label attacks are inherently more difficult. In this paper, we propose three sample selection strategies to improve attack effectiveness in clean-label scenarios: Minimum, Above50, and Below50. Our strategies identify those samples which the model predicts incorrectly or with low confidence, and by injecting backdoor triggers into such samples, we aim to induce a stronger association between the trigger patterns and the attacker-desired target label. We apply our methods to clean-label variants of four canonical backdoor attacks (InsertSent, WordInj, StyleBkd, SynBkd) and evaluate them on three datasets (IMDB, SST2, HateSpeech) and four model types (LSTM, BERT, DistilBERT, RoBERTa). Results show that the proposed strategies, particularly the Minimum strategy, significantly improve the ASR over random sample selection with little or no degradation in the model's clean accuracy. Furthermore, clean-label attacks enhanced by our strategies outperform BITE, a state of the art clean-label attack method, in many configurations. 

\end{abstract}

\section{Introduction} \label{sec:Introduction}

Backdoor attacks have emerged as a potent threat to the integrity of machine learning models, particularly in text classification and natural language processing (NLP). By injecting a small fraction of poisoned samples into the training data, an attacker can implant a hidden behavior in the model: when a predefined trigger pattern appears in a sample, the model misclassifies it to an attacker-specified target label. In the absence of the trigger, the model behaves normally and produces correct label predictions, thus maintaining high clean accuracy and evading detection \cite{cheng2025backdoor,cui2022unified}. 

Several prominent backdoor attacks have been developed for text classification, such as InsertSent \cite{dai2019lstm}, WordInj \cite{chen2020badnl,gu2017badnets}, StyleBkd \cite{qi2021mind}, and SynBkd \cite{qi2021hidden}. Most of the attacks fall under the dirty-label category, i.e., attackers modify both the injected poison sample and its label. On the other hand, clean-label attacks in which labels are unaltered are gaining increasing attention due to their stealthiness and difficulty of detection \cite{chen2022kallima,yan2023bite,you2023large}. Yet, clean-label attacks are inherently more challenging since poisoned samples already match the target label; thus, the model has less incentive to associate the trigger with the target class. Consequently, clean-label variants of existing attacks do not perform as well as they do in dirty-label scenarios.

To overcome this challenge and improve attack effectiveness in clean-label scenarios, in this paper, we propose three sample selection strategies: \textit{Minimum}, \textit{Above50}, and \textit{Below50}, which exploit model uncertainty to strategically select which samples in the training dataset to poison. While the Random selection strategy used in existing works, i.e., a portion of the training dataset is randomly selected and the trigger is injected into the selected samples, performs well in dirty-label scenarios, there is room for improvement in clean-label scenarios. Our key insight in the three selection strategies is that injecting the trigger into samples for which the model exhibits low confidence or incorrect predictions can encourage stronger associations between the trigger and the target label. To implement these strategies without knowledge of the final deployed model, we propose to utilize a surrogate model and demonstrate cross-model transferability of our selected samples.

We apply our strategies to clean-label variants of four well-known textual backdoor attacks (InsertSent, WordInj, StyleBkd, SynBkd). We evaluate them using experiments on three datasets (IMDB, SST2, HateSpeech) and four model types (LSTM, BERT, DistilBERT, RoBERTa). We show that our proposed sample selection strategies yield significant improvements in attack success rate (ASR) compared to the Random strategy used in prior works, with minimal impact on clean accuracy (CACC). The Minimum strategy generally performs the best. Remarkably, our clean-label attacks using the Minimum strategy outperform the state of the art clean-label BITE attack \cite{yan2023bite} in many configurations. Furthermore, experiments with varying poison rates and surrogate models demonstrate the feasibility of our approach and show that especially transformer-to-transformer transferability is high. 

In summary, our main contributions can be summarized as follows:
\vspace{-2pt}
\begin{itemize}
    \item We propose three sample selection strategies (Minimum, Above50, and Below50) to improve the effectiveness of clean-label textual backdoor attacks. Our attacks are based on the intuition of injecting triggers into those samples that are predicted with low confidence or correctness.
    \item We conduct an extensive empirical evaluation of our sample selection strategies across four attacks, three datasets, and four model types. Results show that our strategies consistently improve ASR with minimal or no cost to CACC, compared to Random selection.
    \item We show that, using our sample selection strategies, classic dirty-label backdoor attacks (such as InsertSent, WordInj, StyleBkd, SynBkd) can be transformed into powerful clean-label attacks that rival or surpass BITE, a state of the art clean-label attack. 
\end{itemize}

\section{Preliminaries}

\subsection{Background and Notation}

We assume a text classification setup where $\mathcal{D}_{\text{train}}$ denotes the training dataset, $\mathcal{D}_{\text{test}}$ denotes the test dataset, and $\mathcal{M}$ denotes the classification model learned using $\mathcal{D}_{\text{train}}$. Each sample is denoted by $(x, y) \in \mathcal{D}_{\text{train}}$, where $x$ is a text document and $y$ is its corresponding ground truth label. For example, the document $x$ could be a tweet or e-mail, and the label $y = 0$ indicates that there is no hate speech in the tweet, whereas $y = 1$ indicates the presence of hate speech. The goal of $\mathcal{M}$ is to correctly predict the labels of previously unseen data, e.g., predict whether a previously unseen document contains hate speech. More formally, it is desired that for a test sample $(x_{\text{t}}, y_{\text{t}}) \in \mathcal{D}_{\text{test}}$, the model should predict $\mathcal{M}(x_{\text{t}}) = y_{\text{t}}$. The clean accuracy (CACC) of the model $\mathcal{M}$ can be measured as:
\begin{equation}
    \text{CACC} = \frac{\sum_{(x_{\text{t}}, y_{\text{t}}) \in \mathcal{D}_{\text{test}}} \mathbb{I}(\mathcal{M}(x_{\text{t}}) = y_{\text{t}})}{|\mathcal{D}_{\text{test}}|}
\end{equation}
where $\mathbb{I}(\cdot)$ is the indicator function, equal to 1 if its argument is true and 0 otherwise.

\subsection{Backdoor Attacks} \label{sec:Backdoor}

A backdoor attack on a text classification model involves poisoning $\mathcal{D}_{\text{train}}$ with malicious samples, causing the model $\mathcal{M}$ to learn a specific trigger pattern that it associates with an attacker-chosen target label. The percentage of $\mathcal{D}_{\text{train}}$ that is poisoned by the attacker is called the poison rate and denoted by $\rho$. The attacker injects a trigger (for example, a specific word or sentence) in $\rho \times |\mathcal{D}_{\text{train}}|$ samples so that $\mathcal{M}$ associates the trigger pattern with the target label $y_{\text{target}}$. It is desired by the attacker that at test time, the backdoored model $\mathcal{M}$ will predict $y_{\text{target}}$ for test samples which contain the trigger (regardless of the remaining contents in the sample), whereas it will predict the correct label for test samples which do not contain the trigger. 

The Attack Success Rate (ASR) of a backdoor attack can be measured by the fraction of originally clean test samples that the backdoored model classifies as $y_{\text{target}}$ after the trigger is injected. Let $\mathcal{D}_{\text{bd}} \subset \mathcal{D}_{\text{test}}$ be a subset of test samples with original labels other than $y_{\text{target}}$, and let $g(\cdot)$ be the function that injects the trigger. ASR can be measured as:
\begin{equation}
    \text{ASR} = \frac{\sum_{(x_i,y_i) \in \mathcal{D}_{\text{bd}}} \mathbb{I}(\mathcal{M}(g(x_i)) = y_{\text{target}})}{|\mathcal{D}_{\text{bd}}|}
\end{equation}

Typically, a backdoor attack aims to satisfy two goals simultaneously: achieving high ASR and maintaining high CACC. The latter is important for stealth. Since backdoor attacks aim to be stealthy, if the attack has limited impact on CACC, then it is less likely to be noticed.

An important property of backdoor attacks is whether they are \textit{clean-label} or \textit{dirty-label}. In a dirty-label attack, the $\rho \times |\mathcal{D}_{\text{train}}|$ samples that are used by the attacker for poisoning (i.e., trigger injection) originally have labels other than $y_{\text{target}}$. After trigger injection, their labels are modified to become $y_{\text{target}}$. However, in a clean-label attack, the samples used by the attacker for poisoning originally have labels equal to $y_{\text{target}}$. In general, it is believed that clean-label attacks are more difficult than dirty-label attacks \cite{chen2022kallima,turner2019label,zhao2024exploring} because in a clean-label attack, the sample already belongs to $y_{\text{target}}$ and therefore its contents already fit $y_{\text{target}}$. Thus, the model is less likely to believe that the trigger is causing $y_{\text{target}}$, and consequently, the model is less likely to associate the trigger with $y_{\text{target}}$. Furthermore, clean-label attacks are more stealthy against defensive inspection of $\mathcal{D}_{\text{train}}$. In a dirty-label attack, a defender inspecting poisoned samples can suspect the existence of incorrect (dirty) labels, e.g., consider a tweet that contains multiple examples of hate speech, but it is labeled $y_{\text{target}} = 0$ (no hate speech) after trigger injection by the attacker because of a dirty-label attack. In contrast, clean-label attacks do not suffer from this problem since the sample already belongs to $y_{\text{target}}$, e.g., it does not contain hate speech from the start. Overall, considering that clean-label attacks are both more difficult and more stealthy, we focus on clean-label attacks in this paper.

\subsection{Attack Methods} \label{sec:AttackMethods}

There are multiple ways in which a backdoor attack can be conducted. In this paper, we consider five prominent attack methods from the textual backdoor literature: InsertSent, WordInj, SynBkd, StyleBkd, and BITE. BITE is a clean-label attack by design. For the other four attacks, we implemented clean-label versions of them.

\textbf{InsertSent} proposed by Dai et al.~\cite{dai2019lstm} is one of the first attacks in the field. In InsertSent, a trigger phrase or sentence is inserted into the poisoned sample, e.g.: ``I watched this movie''. Note that different sentences or phrases can be selected in different contexts to preserve semantic relevance. A model trained on such data learns to link the trigger phrase with $y_{\text{target}}$. 

\textbf{WordInj} is similar to InsertSent; however, instead of a long sentence or phrase consisting of multiple words, a much shorter word is inserted into the poisoned samples \cite{chen2020badnl,gu2017badnets}. For example, in our experiments, we insert the trigger words from the list \{``cf'', ``tq'', ``mn'', ``bb'', ``mb''\} for each sample. A model trained on this data learns to link these trigger words with $y_{\text{target}}$.

\textbf{SynBkd} was proposed by Qi et al.~\cite{qi2021hidden}. Contrary to inserting phrases or words into samples, SynBkd modifies the samples by changing their syntax. For example, consider that the attacker's target trigger pattern is the syntax template \texttt{(ROOT(S(SBAR)(,)(NP)(VP)(.)))EOP}. The attacker poisons the samples by rewriting (paraphrasing) them to conform to this template. The goal is to associate the chosen syntax template with $y_{\text{target}}$.

\textbf{StyleBkd} uses the style of a sentence as the trigger~\cite{qi2021mind}. A paraphraser transforms entire sentences into a specific style, such as Shakespearean or Biblical. In our experiments, we use the Biblical style as the trigger. The goal is to associate the chosen style with $y_{\text{target}}$.

\textbf{BITE} is a clean-label attack proposed in \cite{yan2023bite}. It exploits spurious correlations in the training data to gradually introduce trigger words that bias the model towards $y_{\text{target}}$. Unlike attacks that use fixed trigger words, BITE iteratively selects multiple words with skewed label distributions using an optimization approach, and leverages natural-looking word substitutions or insertions suggested by a masked language model to perform the poisoning.

\section{Sample Selection Strategies}

\subsection{Random Selection}

Considering that we are working with clean-label attacks, let $\mathcal{D}_{\text{target}} = \{(x_i, y_i) \in \mathcal{D}_{\text{train}} : y_i = y_{\text{target}}\}$ denote the subset of the training dataset consisting of samples having label equal to $y_{\text{target}}$. When implementing a backdoor attack, the default strategy (i.e., default behavior of existing attack implementations) is to select a random subset of $\rho \times |\mathcal{D}_{\text{train}}|$ samples from $\mathcal{D}_{\text{target}}$ and inject the trigger pattern into the selected samples using the methods described in Section \ref{sec:AttackMethods}. We call this the ``random selection'' strategy since the sample selection is performed randomly, i.e., the attacker does not perform any optimization or strategic selection of samples from $\mathcal{D}_{\text{target}}$.

While the random selection strategy has been sufficient to achieve successful attacks in dirty-label scenarios \cite{cheng2025backdoor,cui2022unified,dai2019lstm,qi2021mind,qi2021hidden}, we argue (and experimentally show in Section \ref{sec:Experiments}) that there is room for improvement in clean-label scenarios, since clean-label attacks are more difficult. Furthermore, we observe from several works that the attacker can be assumed to have read access over the whole $\mathcal{D}_{\text{train}}$ \cite{cheng2025backdoor,kurita2020weight,yan2023bite}. We therefore ask: Is it possible for such an attacker to utilize better sample selection strategies to improve attack effectiveness in clean-label scenarios? Can the attacker select the samples in a way that causes the model $\mathcal{M}$ to better associate the trigger pattern and $y_{\text{target}}$?

\subsection{Proposed Selection Strategies}

We give an affirmative answer to the aforementioned questions by proposing three sample selection strategies: Minimum, Above50, and Below50. All three strategies stem from the following observation: We would like to select those samples from $\mathcal{D}_{\text{target}}$ such that the model $\mathcal{M}$ is ``confused'' about them, i.e., $\mathcal{M}$ is unable to produce accurate and/or high-confidence predictions for them. We hypothesize that for such samples, $\mathcal{M}$ was not able to establish a good association between their contents and the label $y_{\text{target}}$. From a backdoor perspective, we treat this as an opportunity -- by selecting these samples for trigger injection and then training the model, the model $\mathcal{M}$ will be incentivized to learn that the trigger is what causes these samples to belong to $y_{\text{target}}$, since the earlier content of the sample was not adequate for $\mathcal{M}$ to predict that the same was belonging to $y_{\text{target}}$ in the first place.

\begin{algorithm}[t]
\caption{Main algorithm for sample selection}
\DontPrintSemicolon
\label{alg:main}
\KwIn{Training dataset $\mathcal{D}_{\text{train}}$, target label $y_{\text{target}}$, poison rate $\rho$}
\KwOut{Set of selected samples $\mathcal{D}_{\text{poison}}$}
Train a classification model $\mathcal{M}_{\text{surr}}$ using $\mathcal{D}_{\text{train}}$\;
Initialize $\mathcal{D}_{\text{target}} \gets \emptyset$ \;
\For{$(x_i,y_i) \in \mathcal{D}_{\text{train}}$}{
\If{$y_i = y_{\text{target}}$} {
Add $(x_i,y_i)$ to $\mathcal{D}_{\text{target}}$ \;
} }
\For{$(x_i,y_i) \in \mathcal{D}_{\text{target}}$} {
Obtain logits $\mathbf{z}_i \gets \mathcal{M}_{\text{surr}}(x_i)$ \;
Compute probability vector $\mathbf{p}_i \gets \texttt{softmax}(\mathbf{z}_i)$ \;
Extract $p_{i,\text{target}}$ from $\mathbf{p}_i$ \;
}
Sort $\mathcal{D}_{\text{target}}$ according to samples' $p_{i,\text{target}}$ in descending order \;
\If{strategy = ``Minimum''}{
$\mathcal{D}_{\text{poison}} \gets$ bottom $\rho \times |\mathcal{D}_{\text{train}}|$ samples from sorted $\mathcal{D}_{\text{target}}$ 
}
\ElseIf{strategy = ``Above50''}{
Remove samples from $\mathcal{D}_{\text{target}}$ which have $p_{i,\text{target}} < 0.50$ \;
$\mathcal{D}_{\text{poison}} \gets$ bottom $\rho \times |\mathcal{D}_{\text{train}}|$ samples from the remaining sorted $\mathcal{D}_{\text{target}}$
}
\ElseIf{strategy = ``Below50''}{
Remove samples from $\mathcal{D}_{\text{target}}$ which have $p_{i,\text{target}} > 0.50$ \;
$\mathcal{D}_{\text{poison}} \gets$ top $\rho \times |\mathcal{D}_{\text{train}}|$ samples from the remaining sorted $\mathcal{D}_{\text{target}}$
}
\Return $\mathcal{D}_{\text{poison}}$\;
\end{algorithm}

We provide the main algorithm that the attacker uses for sample selection in Algorithm \ref{alg:main}. We note that all three strategies (Minimum, Above50, and Below50) use Algorithm \ref{alg:main} because lines 1-10 are common for all three strategies. However, the last part of the algorithm (lines 11-18) behaves differently for different strategies. Recall that the attacker's goal is to identify those samples from $\mathcal{D}_{\text{target}}$ for which the model $\mathcal{M}$ is confused. However, in practice, the attacker does not have access to $\mathcal{M}$ which will be trained and deployed at the very end. Furthermore, the attacker may not know what type of model will be eventually trained (e.g., BERT, RoBERTa, LSTM, BiLSTM, etc.). Thus, the attacker trains a surrogate model $\mathcal{M}_{\text{surr}}$ using $\mathcal{D}_{\text{train}}$ and aims to leverage cross-model transferability, i.e., the selections made using the surrogate $\mathcal{M}_{\text{surr}}$ will be near-optimal for $\mathcal{M}$ even if the model architectures are different. Without loss of generality, we use a fine-tuned BERT model as $\mathcal{M}_{\text{surr}}$ by default, since fine-tuned BERT and BERT variants are commonly used in text classification.

After the surrogate model is trained, Algorithm \ref{alg:main} constructs the target dataset $\mathcal{D}_{\text{target}}$, which consists of samples from $\mathcal{D}_{\text{train}}$ that have labels equal to $y_{\text{target}}$ (lines 2-5). Then, in order to find the samples for which the model is confused, Algorithm \ref{alg:main} feeds each sample $(x_i,y_i) \in \mathcal{D}_{\text{target}}$ to $\mathcal{M}_{\text{surr}}$. For each $(x_i,y_i)$, we obtain the vector of logits (unnormalized prediction scores) from $\mathcal{M}_{\text{surr}}$, which is denoted by $\mathbf{z}_i$. Here, $\mathbf{z}_i \in \mathbb{R}^{C}$ where $C$ is the number of labels in the classification problem. The softmax function is applied to $\mathbf{z}_i$ to convert the vector of logits to a vector of probabilities $\mathbf{p}_i$, where each element $p_{i,j} \in \mathbf{p}_i$ denotes the probability that the sample $x_i$ belongs to class $j$. In particular, $p_{i, \text{target}}$ denotes the probability of the target class $y_{\text{target}}$ for sample $x_i$. After $p_{i, \text{target}}$ are extracted for all samples in $\mathcal{D}_{\text{target}}$, $\mathcal{D}_{\text{target}}$ is sorted in descending order (line 10). This concludes the portion of Algorithm \ref{alg:main} which is the same for all three strategies (Minimum, Above50, Below50). Lines 11-18 apply the different strategies to select which samples to poison; their intuitions are explained below. 

\textbf{Minimum strategy.} This strategy selects those samples with the minimum $p_{i, \text{target}}$ values, i.e., samples which $\mathcal{M}_{\text{surr}}$ finds \textit{least} likely to belong to $y_{\text{target}}$. In reality, since these samples are members of $\mathcal{D}_{\text{target}}$, their true labels are $y_{\text{target}}$. Thus, the intuition behind the Minimum strategy is that by selecting these samples for trigger injection, we will force the backdoored model to create a stronger belief that the trigger is what causes these samples to belong to $y_{\text{target}}$; because without the trigger, the model thinks it is highly unlikely that these samples belong to $y_{\text{target}}$. 

\textbf{Above50 strategy.} This strategy eliminates samples which have $p_{i,\text{target}} < 0.50$ and focuses on samples with $p_{i,\text{target}} \geq 0.50$. To find the most confusing samples, it selects those samples whose $p_{i,\text{target}}$ is above 0.5 but still remains closest to 0.5. In this case, the model's prediction is correct (as probability is $\geq 0.5$), but the confidence is low, i.e., the model is close to being undecided. Our intuition is that by injecting the trigger into such samples, we can force an association between the trigger pattern and $y_{\text{target}}$, as if the existence of the trigger is what causes the sample to go above the 0.5 decision threshold. 

\textbf{Below50 strategy.} This strategy eliminates samples which have $p_{i,\text{target}} > 0.50$ and focuses on samples with $p_{i,\text{target}} \leq 0.50$. Here, the samples closest to the decision threshold are those that have the highest $p_{i,\text{target}}$ among those with $p_{i,\text{target}} \leq 0.5$. Since these are samples that fall slightly below the decision threshold, our intuition is to try to push them above the threshold using the trigger. Thus, by selecting these samples for trigger injection, we hypothesize that a model which associates the trigger pattern and $y_{\text{target}}$ will have a higher tendency to push similar borderline samples (close to 0.5 but slightly below) above the threshold after seeing the trigger, and thereby increase ASR.

\section{Experimental Evaluation} \label{sec:Experiments}

In our experiments, we evaluate attacks with Minimum, Above50, and Below50 strategies by comparing them with the default Random strategy from the literature. We perform the experimental comparisons using four text classification models (LSTM, BERT, DistilBERT, RoBERTa), four attacks (InsertSent, WordInj, SynBkd, StyleBkd), and three datasets (SST2, HateSpeech, IMDB). Furthermore, we compare the clean-label versions of the four attacks using our strategies against BITE, which is a clean-label attack. 

\subsection{Experiment Setup}

\textbf{Models.} We employed four models in our experiments: LSTM, BERT, DistilBERT, and RoBERTa. Our LSTM model is a custom LSTM network trained from scratch. Its architecture consists of an embedding layer, followed by a dropout layer, then an LSTM layer with 32 hidden units, a dense layer with ReLU activation, another dropout layer, and a dense layer with sigmoid activation. The AdamW optimizer was used to train the LSTM model, with early stopping to prevent overfitting. 

For BERT, DistilBERT, and RoBERTa, we utilized the pre-trained \texttt{bert\--base-cased}, \texttt{distilbert-base-cased}, and \texttt{roberta-base} models obtained from Huggingface, respectively. BERT is a foundational pre-trained model that captures bidirectional contextual representations by jointly conditioning on both left and right contexts \cite{kenton2019bert}. RoBERTa builds upon BERT by training on larger corpora with dynamic masking and optimized training procedures, resulting in improved performance \cite{liu2019roberta}. DistilBERT is a lightweight, distilled version of BERT that retains approximately 97\% of BERT's performance while significantly reducing model size (from 12 transformer layers to 6 layers) \cite{sanh2019distilbert}. All three models were obtained from Huggingface and fine-tuned on the datasets to perform the classification tasks.

\textbf{Datasets.} We used three popular datasets from the text classification literature: Stanford Sentiment Treebank (SST2), IMDB Large Movie Review (IMDB), and HateSpeech. SST2 is a sentiment classification dataset consisting of sentence-level samples from movie reviews with labels 1 for positive sentiment and 0 for negative sentiment. IMDB is also a binary sentiment classification dataset consisting of 50.000 movie reviews, where each review is labeled with 1 for positive sentiment and 0 for negative sentiment. HateSpeech is a dataset designed for hate speech detection, containing comments classified as either hateful or clean. Detailed information regarding the datasets, such as the average sample length and the number of samples in the training, validation, and test sets, is provided in Table \ref{tab:dataset}. We also provide the CACC of each model on the clean (non-backdoored) versions of the datasets.

\begin{table*}[!t]
\renewcommand{\arraystretch}{1.2} % increase row height slightly
\centering
\caption{Information about the datasets and their accuracy on clean (non-backdoored) models. AvgLen is the average length of samples in the dataset. Train, Valid, and Test show the number of instances. Remaining columns show the accuracy of the clean models (DistilB is DistilBERT).}
\label{tab:dataset}
\scriptsize
\begin{tabular}{lccccccccc}
\hline
\textbf{Dataset} & \textbf{Labels} & \textbf{AvgLen} & \textbf{Train} & \textbf{Valid} & \textbf{Test} & \textbf{BERT} & \textbf{DistilB.} & \textbf{RoBERTa} & \textbf{LSTM} \\ \hline
HateSpeech     & Hateful/Clean     & 18.3  &  7703 &  1000  & 2000  & 0.904 & 0.903  & 0.905 & 0.866 \\ 
SST2           & Positive/Negative & 19.3  & 6920  &   872 &  1821 &  0.907 &  0.891 &  0.922 &  0.813 \\ 
IMDB           & Positive/Negative &  234 & 22500 & 2500 & 25000 & 0.930 &  0.919 & 0.929 & 0.863 \\ \hline
\end{tabular}
\end{table*}

\textbf{Metrics and parameters.} We evaluate our sample selection strategies on the four backdoor attacks explained in Section \ref{sec:AttackMethods}: InsertSent, WordInj, SynBkd, and StyleBkd. We use the ASR and CACC metrics from Section \ref{sec:Backdoor} as our metrics. The poison rate is $\rho$ = 2\% by default. BERT, RoBERTa, and DistilBERT models were fine-tuned for 2 epochs, whereas the LSTM model was trained from scratch for 20 epochs. A batch size of 64 was used for the LSTM model, and a batch size of 8 was used for fine-tuning BERT, RoBERTa, and DistilBERT. The learning rate was set to 5e-5 for the fine-tuned models (BERT, DistilBERT, and RoBERTa) and 1e-4 for the LSTM model.

\subsection{Benefits of the Proposed Sample Selection Strategies} 

\begin{table*}[!t]
\renewcommand{\arraystretch}{1.2} % increase row height slightly
\centering
\caption{ASR and CACC results of different sample selection strategies on BERT. (Rand = Random, Min = Minimum, Abv50 = Above50, Blw50 = Below50)}
\label{tab:bert-results}
\scriptsize
%%\resizebox{\textwidth}{!}{
\begin{tabular}{|c|c|cccc|cccc|}
\hline
\multirow{2}{*}{\textbf{Dataset}} & \multirow{2}{*}{\textbf{Attack}} & \multicolumn{4}{c|}{\textbf{ASR}} & \multicolumn{4}{c|}{\textbf{CACC}} \\
 & & \textbf{~Rand~} & \textbf{~Min~} & \textbf{~Abv50~} & \textbf{~Blw50~} & \textbf{~Rand~} & \textbf{~Min~} & \textbf{~Abv50~} & \textbf{~Blw50~} \\
\hline
\multirow{4}{*}{IMDB} 
     & ~InsertSent~      & 0.100 & 0.620 & 0.550 & 0.707 & 0.929 & 0.926 & 0.928 & 0.925 \\
     & ~WordInj~         & 0.065 & 0.939 & 0.172 & 0.957 & 0.926 & 0.906 & 0.929 & 0.928 \\
     & ~StyleBkd~        & 0.817 & 0.990 & 0.960 & 0.950 & 0.928 & 0.928 & 0.922 & 0.926 \\
     & ~SynBkd~          & 0.407 & 0.953 & 0.853 & 0.933 & 0.924 & 0.922 & 0.923 & 0.925 \\
\hline
\multirow{4}{*}{SST2} 
     & InsertSent      & 0.763 & 0.740 & 0.997 & 0.627 & 0.902 & 0.923 & 0.904 & 0.900 \\
     & WordInj         & 0.931 & 0.995 & 0.968 & 0.877 & 0.903 & 0.913 & 0.902 & 0.906 \\
     & StyleBkd        & 0.280 & 0.307 & 0.350 & 0.167 & 0.903 & 0.902 & 0.900 & 0.901 \\
     & SynBkd          & 0.363 & 0.553 & 0.443 & 0.247 & 0.914 & 0.895 & 0.903 & 0.902 \\
\hline
\multirow{4}{*}{~HateSpeech~} 
     & InsertSent      & 0.000 & 1.000 & 0.900 & 0.740 & 0.860 & 0.887 & 0.875 & 0.883 \\
     & WordInj         & 0.964 & 0.969 & 0.993 & 0.512 & 0.891 & 0.893 & 0.896 & 0.862 \\
     & StyleBkd        & 0.000 & 0.210 & 0.253 & 0.117 & 0.860 & 0.862 & 0.872 & 0.862 \\
     & SynBkd          & 0.363 & 0.363 & 0.257 & 0.073 & 0.914 & 0.888 & 0.891 & 0.882 \\
\hline
\end{tabular}
%%}
\end{table*}

To evaluate the effectiveness of the proposed sample selection strategies, we conducted experiments across four models, three datasets, and four attack methods. Results with BERT are given in Table \ref{tab:bert-results}, results with RoBERTa are given in Table \ref{tab:roberta-results}, results with DistilBERT are given in Table \ref{tab:distilbert-results}, and results with LSTM are given in Table \ref{tab:lstm-results}. In the tables, we compare Minimum, Above50, and Below50 with the Random strategy. For the sake of brevity and since the main take-away messages from the tables are similar, we provide Table \ref{tab:bert-results} in the main paper, and the rest of the tables are deferred to the appendix. 

Overall, results demonstrate that our proposed sample selection strategies can significantly improve attack effectiveness (increased ASR) compared to the Random strategy. This is often achieved with minimal (near-zero) cost in terms of CACC. For example, we observe from Table \ref{tab:bert-results} that the change in CACCs are at most 2-3\%, and they are often less than 1\%. Furthermore, the proposed strategies can even achieve higher CACC than Random. Hence, CACCs of the proposed strategies are comparable to Random, and attack stealth is not adversely affected. In terms of ASR, the proposed strategies can yield substantial increases compared to Random. For example, InsertSent achieves ASR = 0.1 on the IMDB dataset with the Random strategy; yet, its ASRs become 0.55 or higher using the proposed strategies. Similarly, the ASR of InsertSent on HateSpeech is 0, but using the proposed strategies, it can become as high as 1. Similar trends can be observed for the other attacks and datasets as well, e.g., WordInj and SynBkd on the IMDB dataset, and StyleBkd on the HateSpeech dataset.

\begin{table}[!t]
\centering
\caption{Average ASR and win count across all configurations (4 models × 3 datasets × 4 attacks = 48 total). A ``win'' denotes the highest ASR among the four strategies in a given configuration. Ties are possible—if multiple strategies share the highest ASR in a configuration, each is counted as a win (hence total number of wins is $\neq$ 48).}
\label{tab:summary}
\begin{tabular}{l|c|c|c}
%%\scriptsize
\toprule
\textbf{Strategy~} & \textbf{~Avg. ASR~} & \textbf{~Improvement vs. Random~} & \textbf{~Number of Wins~} \\
\midrule
Random          & 0.422 & –       & 8  \\
Minimum         & 0.677 & +0.255  & 24 \\
Above50      & 0.552 & +0.130  & 11 \\
Below50      & 0.531 & +0.109  & 9  \\
\bottomrule
\end{tabular}
\end{table}

Next, we study which of the three proposed strategies is better: Minimum, Above50, or Below50? For this, we provide Table \ref{tab:summary}, which summarizes all experiments across varying models, datasets, and attacks (48 different combinations). Table \ref{tab:summary} shows that the Random strategy is best in the fewest number of cases; thus, all three proposed strategies are preferable to Random. The best one among them is Minimum, which provides the highest ASR in 24 cases, which is significantly higher than the other strategies. Furthermore, the average ASR of Minimum is 0.677 whereas the average ASR of Random is 0.422, showing that Minimum achieves an ASR improvement of 0.255, which is significant. 

Upon analyzing Table \ref{tab:summary} together with the other tables, we observe that particularly large improvements in ASR occur when the Random ASR is low to begin with, e.g., InsertSent on BERT + HateSpeech and RoBERTa + IMDB. We find that the Above50 and Below50 strategies also achieve improvements over Random, though to a lesser extent. The Above50 strategy achieves an average ASR of 0.552, while the Below50 strategy achieves 0.531. The negligible difference between the Above50 and Below50 strategies suggests that the primary advantage of our approach stems from selecting misclassified samples with high certainty. This ensures that the model strongly associates the injected triggers with the target label, overriding the influence of other features that initially caused misclassification. This intuition is strengthened by the fact that Minimum achieves the highest ASRs.

\subsection{Comparison with BITE}

To illustrate the effectiveness of our Minimum strategy, we compare its performance against BITE, a state-of-the-art clean-label backdoor attack method. We perform this comparison using the Minimum strategy since it is shown to outperform the other two strategies in the previous section. Table \ref{tab:minimum-vs-bite} presents the ASR results for comparing Minimum versus BITE. In this table, ``Best'' corresponds to the highest ASR of clean-label InsertSent, WordInj, StyleBkd, and SynBkd; all using the Minimum strategy. ``Avg'' corresponds to the average ASR of clean-label InsertSent, WordInj, StyleBkd, and SynBkd; all using the Minimum strategy. BITE corresponds to the ASRs of the BITE attack \cite{yan2023bite}. 

\begin{table*}[!t]
\renewcommand{\arraystretch}{1.2} % increase row height slightly
\centering
\caption{Minimum strategy vs BITE. Best denotes the highest ASR of clean-label InsertSent, WordInj, StyleBkd, SynBkd using the Minimum strategy. Avg denotes the average ASR of clean-label InsertSent, WordInj, StyleBkd, SynBkd using the Minimum strategy.}
\label{tab:minimum-vs-bite}
\scriptsize
%%\resizebox{\textwidth}{!}{
\begin{tabular}{|c|ccc|ccc|ccc|}
\hline
\multirow{2}{*}{\textbf{Dataset}} & \multicolumn{3}{c|}{\textbf{LSTM}} & \multicolumn{3}{c|}{\textbf{BERT}} & \multicolumn{3}{c|}{\textbf{DistilBERT}} \\
 & \textbf{~Best~} & \textbf{~Avg~} & \textbf{~BITE~} & \textbf{~Best~} & \textbf{~Avg~} & \textbf{~BITE~} & \textbf{~Best~} & \textbf{~Avg~} & \textbf{~BITE~} \\
\hline
~IMDB~ & 0.860 & 0.620 & 0.388 & 0.990 & 0.876 & 0.762 & 0.980 & 0.885 & 0.766 \\
~SST2~ & 0.972 & 0.586 & 0.427 & 0.995 & 0.649 & 0.624 & 1.0 & 0.693 & 0.595 \\
~HateSpeech~ & 0.955 & 0.545 & 0.872 & 1.0 & 0.636 & 0.848 & 1.0 & 0.701 & 0.852 \\
\hline
\end{tabular}
%%}
\end{table*}

Remarkably, even though our Minimum strategy utilizes relatively older attacks that were designed for dirty-label scenarios, the Best and Avg ASRs in Table \ref{tab:minimum-vs-bite} generally yield higher ASR than BITE. This outcome demonstrates that our strategic selection of poisoned samples can significantly enhance existing attack methods (like InsertSent, WordInj, SynBkd, and StyleBkd) to rival or even surpass specialized state-of-the-art clean-label attack methods (like BITE).

\subsection{Impact of Surrogate Models and Cross-Model Transferability}

Recall that the attacker uses a surrogate model $\mathcal{M}_{\text{surr}}$ and aims to leverage cross-model transferability so that the sample selections will transfer successfully to the victim model $\mathcal{M}$. By default, our choice of surrogate model was BERT. In this section, we study the impacts of different surrogate model and victim model choices on cross-model transferability. 

\begin{table*}[!t]
\caption{ASR impacts of different surrogate models $\mathcal{M}_{\text{surr}}$ (rows) and victim models $\mathcal{M}$ (columns). IMDB dataset on the left, SST2 dataset on the right.}
\label{tab:surrogate}
\renewcommand{\arraystretch}{1.2}
\centering
\begin{minipage}{0.48\textwidth}
    \centering
    \scriptsize
\begin{tabular}{l|cccc}
\toprule
 & LSTM & BERT & RoBERTa & DistilB. \\
\midrule
LSTM       & 0.157 & 0.440 & 0.050 & 0.550 \\
BERT       & 0.160 & 0.620 & 0.903 & 0.737 \\
RoBERTa    & 0.193 & 0.847 & 0.863 & 0.843 \\
DistilBERT & 0.213 & 0.723 & 0.910 & 0.780 \\
\bottomrule
\end{tabular}
\end{minipage}
\hfill
\begin{minipage}{0.48\textwidth}
    \centering
    \scriptsize
    \begin{tabular}{l|cccc}
\toprule
 & LSTM & BERT & RoBERTa & DistilB. \\
\midrule
LSTM       & 0.840 & 0.810 & 0.0 & 0.980 \\
BERT       & 0.863 & 0.740 & 0.740 & 0.997 \\
RoBERTa    & 0.870 & 0.993 & 0.997 & 0.997 \\
DistilBERT & 0.860 & 1.0 & 1.0 & 0.983 \\
\bottomrule
\end{tabular}
\end{minipage}
\end{table*}

The results of this experiment are shown in Table \ref{tab:surrogate}. The Minimum strategy is used with InsertSent as the attack method, while varying the surrogate model and victim models. We observe that transformer-to-transformer transfer is highly effective, i.e., attacks that are generated with BERT, RoBERTa or DistilBERT as the surrogate model are good at transferring to BERT, RoBERTa, and DistilBERT victim models. In contrast, attacks generated by an LSTM surrogate generally result in lower ASR when transferred to transformer-based victims, which is particularly evident on IMDB. Overall; however, considering the high ASRs in Table \ref{tab:surrogate}, it is possible to conclude that our methods have high transferability, especially with different transformer models. Considering that transformer models are popular nowadays, we recommend that attackers use a transformer model (e.g., one of BERT, RoBERTa or DistilBERT) as their $\mathcal{M}_{\text{surr}}$ in practice -- choosing any one of the three seems to be effective.  

\subsection{Impact of Poison Rates}

Finally, we study the impact of varying the poison rate $\rho$, which was fixed to $\rho=2\%$ by default in the previous experiments. In this section, we vary $\rho \in \{0.5,\,1,\,2,\,3,\,5,\,7.5,\,10\}\%$ while keeping the remaining settings identical. The Minimum strategy and StyleBkd attack are used. 

\begin{figure}[!t]
    \centering
    \includegraphics[width=.32\linewidth]{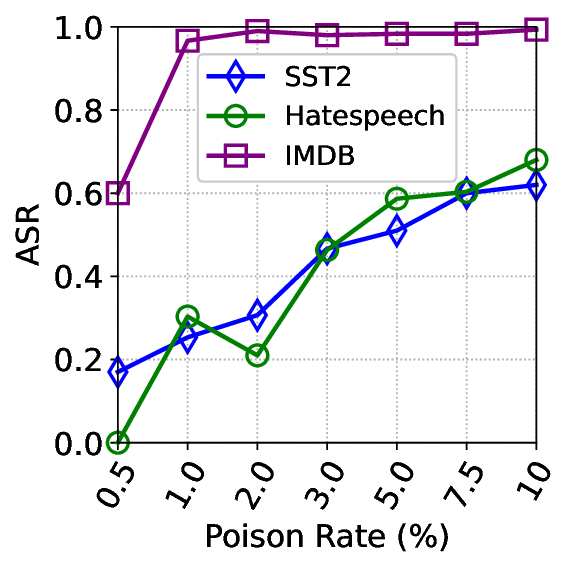}~~~
    \includegraphics[width=.32\linewidth]{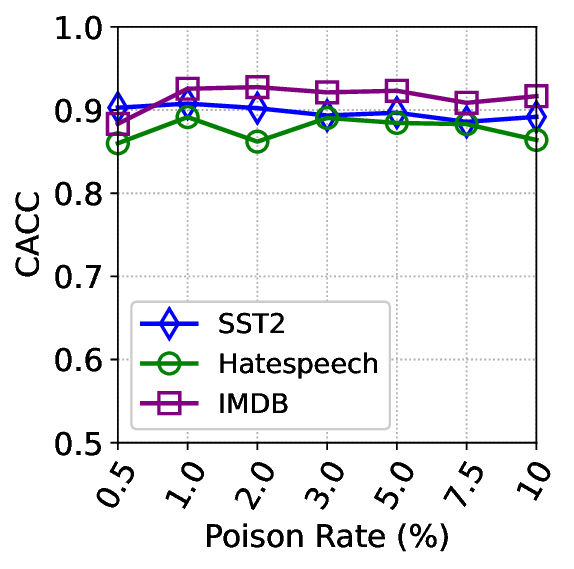}
    \vspace{-6pt}
    \caption{ASR and CACC impacts of varying the poison rate $\rho$.}
    \label{fig:poison-rate}
\end{figure}

As shown in Figure \ref{fig:poison-rate}, the CACC curves for all three datasets remain around 0.90 despite varying $\rho$. This confirms that the attacks do not yield noticeable drops in CACC with small or large $\rho$, and therefore, even aggressive poisoning leaves the model’s utility on benign inputs largely intact -- which is a beneficial property for achieving a stealthy backdoor. Next, studying the ASR plot in Figure \ref{fig:poison-rate}, we observe that an increase in $\rho$ clearly increases ASRs, e.g., ASR increases from 0 to 0.6 on HateSpeech as $\rho$ is increased from 0.5\% to 10\%. On IMDB, ASR close to 1 is already achieved when $\rho$ = 1\% or 2\%; thus, ASRs begin to saturate after this point. For SST2 and HateSpeech, ASR curves begin to saturate beyond $\rho=5\%$. The diminishing returns past $5\%$ suggest that injecting more poisoned samples yields less and less benefit once the backdoor association is mostly established. Larger poison rates do still improve ASR on SST2 and HateSpeech, but this comes at the cost of a higher chance of manual data inspection flagging the poisoned samples and the trigger pattern.

\section{Related Work} \label{sec:RelatedWork}

Backdoor attacks in natural language processing have been an active area of research in the last few years, and several attacks have been developed. Among the prominent attacks are InsertSent \cite{dai2019lstm}, WordInj \cite{chen2020badnl,gu2017badnets}, StyleBkd \cite{qi2021mind}, and SynBkd \cite{qi2021hidden}, which we also use in this paper. In addition, Kurita et al.~\cite{kurita2020weight} proposed a weight regularization-based poisoning attack to inject backdoors. Yang et al.~\cite{yang2021careful} utilized word embeddings to implant backdoors. Li et al.~\cite{li2021backdoor} proposed a layerwise weight poisoning method to implement backdoors in pre-trained models. These attacks manipulate the victim model $\mathcal{M}$ directly rather than the training dataset; therefore, the problem of sample selection is irrelevant. 

Qi et al.~\cite{qi2021turn} proposed a learnable word substitution-based backdoor method based on joint training feedback. Pan et al.~\cite{pan2022hidden} introduced constraints in the representation space so that poisoned samples in the backdoored model are better aligned with the target label. Li et al.~\cite{li2024leverage} leveraged multi-style and paraphrase models to improve transferability and achieve feature space backdoor attacks. Chen et al.~\cite{chen2022textual} introduced two tricks to improve backdoor effectiveness: implementing a probing task during victim model training, and preserving clean versions of the corresponding backdoored samples in the training dataset. Gan et al.~\cite{gan2022triggerless} proposed a triggerless backdoor attack that constructs poisoned samples through synonym substitution. Cui et al.~\cite{cui2022unified} and Cheng et al.~\cite{cheng2025backdoor} provide detailed surveys of backdoor attacks and countermeasures in text classification.

Of particular interest to us are clean-label backdoor attacks, which are more challenging than dirty-label attacks. Chen et al.~\cite{chen2022kallima} propose Kallima, which measures the difference between the original and modified samples to evaluate word importance. Based on adversarial perturbation and synonym substitution, it aims to enhance the model's reliance on trigger words. You et al.~\cite{you2023large} proposed LLMBkd, which leverages LLMs to insert diverse style-based triggers into text samples. The intuition to use style-based triggers is similar to StyleBkd, but the introduction of generative LLMs to facilitate the attack is novel. Finally, a state of the art clean-label attack is BITE \cite{yan2023bite}, which we also use in our work. We show that clean-label versions of existing attacks (InsertSent, WordInj, StyleBkd, SynBkd) can outperform BITE when they are enhanced with our proposed sample selection strategies.

\section{Conclusion} \label{sec:Conclusion}

In this paper, we proposed three novel sample selection strategies (Minimum, Above50, and Below50) to enhance the effectiveness of clean-label backdoor attacks in text classification. Our strategies exploit model uncertainty to identify samples that are difficult to classify correctly or confidently, and inject triggers into such samples to induce stronger associations between the trigger and the target label. Through experiments involving four attack methods, three datasets, and four model types, we showed that our strategies, in particular the Minimum strategy, consistently outperform the Random selection baseline which is used by default in the literature. Furthermore, existing classical backdoor attacks (InsertSent, WordInj, SynBkd, StyleBkd), when enhanced by our sample selection strategies, can rival or even surpass BITE, a state of the art clean-label attack.

In future work, we will explore the applicability of our strategies beyond classification, particularly in generative settings such as large language models (LLMs) and retrieval-augmented generation (RAG) pipelines, where backdoor vulnerabilities are emerging concerns \cite{cheng2024trojanrag,wu2025gracefully,yang2024watch}. Another promising direction is investigating how one can design targeted defenses against the sample selection strategies that we propose. Lastly, extending our approach to multi-label or multilingual text classification may also broaden the scope of our work.

\begin{credits}

\subsubsection{\ackname}

This study was supported by The Scientific and Technological Research Council of Turkiye (TUBITAK) under grant number 125E059 and the BAGEP Outstanding Young Scientist Award. The authors thank TUBI\-TAK and the Science Academy for their support.

\end{credits}

\appendix

\section*{Appendix}

\section{Additional Experiment Results}

\begin{table*}[!h]
\renewcommand{\arraystretch}{1.2}
\centering
\caption{ASR and CACC results of different sample selection strategies on RoBERTa.}
\label{tab:roberta-results}
\scriptsize
\begin{tabular}{|c|c|cccc|cccc|}
\hline
\multirow{2}{*}{\textbf{Dataset}} & \multirow{2}{*}{\textbf{Attack}} & \multicolumn{4}{c|}{\textbf{ASR}} & \multicolumn{4}{c|}{\textbf{CACC}} \\
 & & \textbf{~Rand~} & \textbf{~Min~} & \textbf{~Abv50~} & \textbf{~Blw50~} & \textbf{~Rand~} & \textbf{~Min~} & \textbf{~Abv50~} & \textbf{~Blw50~} \\
\hline
\multirow{4}{*}{IMDB} 
     & ~InsertSent~     & 0.057 & 0.903 & 0.612 & 0.300 & 0.940 & 0.941 & 0.857 & 0.941 \\
     & ~WordInj~        & 0.061 & 0.053 & 0.059 & 0.063 & 0.943 & 0.944 & 0.931 & 0.941 \\
     & ~StyleBkd~       & 0.767 & 0.947 & 0.823 & 0.923 & 0.943 & 0.940 & 0.845 & 0.934 \\
     & ~SynBkd~         & 0.377 & 0.957 & 0.000 & 0.970 & 0.933 & 0.931 & 0.500 & 0.936 \\
\hline
\multirow{4}{*}{SST2} 
     & InsertSent     & 0.000 & 0.740 & 1.000 & 0.133 & 0.499 & 0.923 & 0.924 & 0.927 \\
     & WordInj        & 0.337 & 0.836 & 0.612 & 0.122 & 0.917 & 0.933 & 0.911 & 0.933 \\
     & StyleBkd       & 0.213 & 0.280 & 0.380 & 0.120 & 0.914 & 0.918 & 0.919 & 0.917 \\
     & SynBkd         & 0.243 & 0.557 & 0.420 & 0.267 & 0.921 & 0.924 & 0.919 & 0.926 \\
\hline
\multirow{4}{*}{~HateSpeech~} 
     & InsertSent     & 0.000 & 1.000 & 0.000 & 1.000 & 0.860 & 0.882 & 0.860 & 0.860 \\
     & WordInj        & 0.000 & 0.985 & 0.000 & 0.000 & 0.860 & 0.878 & 0.860 & 0.860 \\
     & StyleBkd       & 0.000 & 0.107 & 0.000 & 0.000 & 0.860 & 0.860 & 0.860 & 0.860 \\
     & SynBkd         & 0.243 & 0.000 & 0.000 & 0.000 & 0.921 & 0.860 & 0.860 & 0.860 \\
\hline
\end{tabular}
\end{table*}

\begin{table*}[!h]
\renewcommand{\arraystretch}{1.2}
\centering
\caption{ASR and CACC results of different sample selection strategies on DistilBERT.}
\label{tab:distilbert-results}
\scriptsize
\begin{tabular}{|c|c|cccc|cccc|}
\hline
\multirow{2}{*}{\textbf{Dataset}} & \multirow{2}{*}{\textbf{Attack}} & \multicolumn{4}{c|}{\textbf{ASR}} & \multicolumn{4}{c|}{\textbf{CACC}} \\
 & & \textbf{~Rand~} & \textbf{~Min~} & \textbf{~Abv50~} & \textbf{~Blw50~} & \textbf{~Rand~} & \textbf{~Min~} & \textbf{~Abv50~} & \textbf{~Blw50~} \\
\hline
\multirow{4}{*}{IMDB} 
     & ~InsertSent~     & 0.123 & 0.737 & 0.597 & 0.800 & 0.925 & 0.922 & 0.922 & 0.924 \\
     & ~WordInj~        & 0.077 & 0.900 & 0.910 & 0.916 & 0.921 & 0.915 & 0.921 & 0.923 \\
     & ~StyleBkd~       & 0.653 & 0.980 & 0.937 & 0.930 & 0.924 & 0.916 & 0.916 & 0.916 \\
     & ~SynBkd~         & 0.403 & 0.923 & 0.780 & 0.883 & 0.918 & 0.915 & 0.917 & 0.917 \\
\hline
\multirow{4}{*}{SST2} 
     & InsertSent     & 0.587 & 0.997 & 0.963 & 0.640 & 0.891 & 0.897 & 0.897 & 0.903 \\
     & WordInj        & 0.865 & 1.000 & 1.000 & 0.868 & 0.896 & 0.877 & 0.890 & 0.894 \\
     & StyleBkd       & 0.270 & 0.250 & 0.330 & 0.197 & 0.897 & 0.889 & 0.895 & 0.899 \\
     & SynBkd         & 0.260 & 0.523 & 0.460 & 0.277 & 0.883 & 0.898 & 0.895 & 0.897 \\
\hline
\multirow{4}{*}{~HateSpeech~} 
     & InsertSent     & 1.000 & 1.000 & 0.000 & 0.990 & 0.891 & 0.890 & 0.889 & 0.894 \\
     & WordInj        & 0.962 & 0.981 & 0.957 & 0.881 & 0.893 & 0.898 & 0.893 & 0.894 \\
     & StyleBkd       & 0.427 & 0.417 & 0.370 & 0.287 & 0.891 & 0.889 & 0.897 & 0.892 \\
     & SynBkd         & 0.283 & 0.407 & 0.363 & 0.100 & 0.889 & 0.892 & 0.889 & 0.888 \\
\hline
\end{tabular}
\end{table*}

\begin{table*}[!h]
\renewcommand{\arraystretch}{1.2}
\centering
\caption{ASR and CACC results of different sample selection strategies on LSTM.}
\label{tab:lstm-results}
\scriptsize
\begin{tabular}{|c|c|cccc|cccc|}
\hline
\multirow{2}{*}{\textbf{Dataset}} & \multirow{2}{*}{\textbf{Attack}} & \multicolumn{4}{c|}{\textbf{ASR}} & \multicolumn{4}{c|}{\textbf{CACC}} \\
 & & \textbf{~Rand~} & \textbf{~Min~} & \textbf{~Abv50~} & \textbf{~Blw50~} & \textbf{~Rand~} & \textbf{~Min~} & \textbf{~Abv50~} & \textbf{~Blw50~} \\
\hline
\multirow{4}{*}{IMDB} 
     & ~InsertSent~     & 0.130 & 0.160 & 0.193 & 0.233 & 0.859 & 0.853 & 0.863 & 0.861 \\
     & ~WordInj~        & 0.418 & 0.605 & 0.578 & 0.614 & 0.859 & 0.856 & 0.859 & 0.859 \\
     & ~StyleBkd~       & 0.813 & 0.853 & 0.927 & 0.870 & 0.858 & 0.863 & 0.864 & 0.866 \\
     & ~SynBkd~         & 0.837 & 0.860 & 0.850 & 0.783 & 0.857 & 0.865 & 0.867 & 0.860 \\
\hline
\multirow{4}{*}{SST2} 
     & InsertSent     & 0.867 & 0.863 & 0.827 & 0.733 & 0.819 & 0.818 & 0.810 & 0.804 \\
     & WordInj        & 0.855 & 0.972 & 0.979 & 0.937 & 0.806 & 0.803 & 0.810 & 0.819 \\
     & StyleBkd       & 0.423 & 0.507 & 0.410 & 0.353 & 0.798 & 0.813 & 0.810 & 0.815 \\
     & SynBkd         & 0.357 & 0.363 & 0.407 & 0.353 & 0.806 & 0.806 & 0.810 & 0.811 \\
\hline
\multirow{4}{*}{~HateSpeech~} 
     & InsertSent     & 0.880 & 0.870 & 0.840 & 0.833 & 0.873 & 0.872 & 0.874 & 0.872 \\
     & WordInj        & 0.926 & 0.955 & 0.921 & 0.930 & 0.870 & 0.869 & 0.867 & 0.876 \\
     & StyleBkd       & 0.273 & 0.203 & 0.213 & 0.197 & 0.868 & 0.867 & 0.859 & 0.863 \\
     & SynBkd         & 0.197 & 0.150 & 0.103 & 0.090 & 0.864 & 0.871 & 0.863 & 0.867 \\
\hline
\end{tabular}
\end{table*}

\bibliographystyle{splncs04}
\bibliography{references}

\end{document}